\PassOptionsToPackage{table,xcdraw}{xcolor}

\documentclass[acmtog,nonacm]{acmart}

\AtBeginDocument{%
  }

\usepackage{tabularx}
\usepackage{booktabs}
\usepackage{wrapfig}
\usepackage{algorithm}
\usepackage{pifont}
\usepackage{algpseudocode}
\usepackage{subcaption}
\usepackage{graphicx}
\usepackage{makecell}
\usepackage{tikz}
\usetikzlibrary{patterns}

\newcommand{\cmark}{\ding{51}}%
\newcommand{\xmark}{\ding{55}}%

\begin{document}

\title[Topology- and Geometry-Exact Coupling for Incompressible Fluids and Thin Deformables]
{Topology- and Geometry-Exact Coupling\\for Incompressible Fluids and Thin Deformables} 

\author{Jonathan Panuelos}
\email{jonathan.panuelos@mail.utoronto.ca}
\affiliation{%
  \institution{University of Toronto}
  \city{Toronto}
  \state{Ontario}
  \country{Canada}
}

\author{Eitan Grinspun}
\affiliation{%
  \institution{University of Toronto}
  \city{Toronto}
  \state{Ontario}
  \country{Canada}
}

\author{David Levin}
\affiliation{%
  \institution{University of Toronto}
  \city{Toronto}
  \state{Ontario}
  \country{Canada}
}

\renewcommand{\shortauthors}{Panuelos et al.}

\begin{abstract}
We introduce a topology-preserving discretization for coupling incompressible fluids with thin deformable structures, achieving guaranteed leakproofness through preservation of fluid domain connectivity.
Our approach leverages a stitching algorithm applied to a clipped Voronoi diagram generated from Lagrangian fluid particles, in order to maintain path connectivity around obstacles.
This geometric discretization naturally conforms to arbitrarily thin structures, enabling boundary conditions to be enforced exactly at fluid-solid interfaces.
By discretizing the pressure projection equations on this conforming mesh, we can enforce velocity boundary conditions at the interface for the fluid while applying pressure forces directly on the solid boundary, enabling sharp two-way coupling between phases.
The resulting method prevents fluid leakage through solids while permitting flow wherever a continuous path exists through the fluid domain.
We demonstrate the effectiveness of our approach on diverse scenarios including flows around thin membranes, complex geometries with narrow passages, and deformable structures immersed in liquid, showcasing robust two-way coupling without artificial sealing or leakage artifacts.
\end{abstract}


\begin{teaserfigure}
    \centering
    \includegraphics[width=\textwidth]{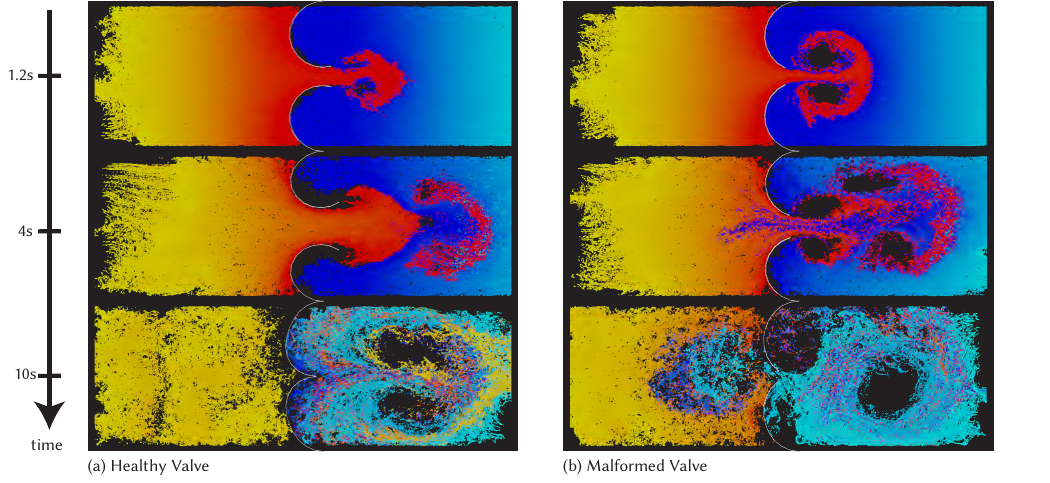}
    \caption{\textbf{Flexible heart valve with leakproof fluid–structure coupling through multiple pump cycles.} Frames 60, 120, and 180 are shown from top to bottom, with fluid being coloured by their starting positions (yellow-red on the left and blue-cyan on the right). Flow through a deformable valve inside a tube is driven by a cyclical body force. (a) In a properly resolved healthy valve, forward flow opens the valve and allows fluid to pass, while reverse flow closes the valve and creates a leakproof separation between the two sides. (b) Our method can distinguish flow between a healthy valve and a malformed valve with annular dilation, an enlargement of the valve opening leading to fluid regurgitation, as shown by the significant amount of blue fluid in the left chamber.}
    \label{fig:teaser}
\end{teaserfigure}
\maketitle

\section{Introduction}
Heart valves directing blood flow, parachutes slowing falling payloads, and fans blowing air are widely varying phenomena the modelling of which crucially depend on the interaction between incompressible fluid flow and codimensional geometry.
Existing methods for coupling incompressible fluids with deformable structures face fundamental limitations when dealing with thin or codimensional features.
Approaches that use separate discretizations for fluid and solid domains suffer from geometric misalignment at interfaces, causing numerical leakage where fluid inappropriately flows through solid barriers.

Conversely, volumetric representations of thin structures introduce artificial thickness that can seal off valid flow paths, preventing fluid from passing through narrow orifices and gaps.
These issues become particularly problematic in scenarios involving large deformations, where topology changes in the fluid phase become common.
Consider the rhythmic opening and closing of a heart valve for example---leaky discretizations cause incorrect backflow while the opposite restricts flow inaccurately.

Addressing these challenges requires a discretization that is leak\-proof only where required, preventing flow through solids while preserving all valid fluid paths.
The key requirement is that the fluid domain’s \emph{path connectivity} must exactly match that induced by the solid geometry (``topologically exact''). This is realized through a flux-form finite-volume discretization on a modified unstructured Voronoi diagram.
Solid boundaries are explicitly included as finite-volume faces, rather than approximated or sampled separately (``geometrically exact'').
As a consequence, incompressibility enforces exact volumetric flux balance over any connected fluid region, enabling geometrically consistent volumetric coupling even through narrow fluid passages and around immersed codimensional solid structures.

To achieve topological and geometric exactness, we extend Voronoi-based Lagrangian fluid methods through a \emph{clipped Voronoi stitching algorithm}.
Our approach constructs a Voronoi diagram from Lagrangian fluid particles and then clips it to include solid boundary faces as explicit interfaces in the mesh structure.
Cells newly created by the clipping process that do not enclose their generating particle, which we denote as ``orphaned cells'', are reassigned to neighboring fluid particles through a connectivity-preserving procedure that maintains the fluid domain’s path structure.
This ensures that the resulting partition conforms exactly to thin solid structures while preserving all flowable regions.

The resulting conforming mesh supports discretization of the pressure projection equations, with the Poisson solve incorporating contributions from both fluid–fluid and fluid–solid interfaces.
Coupling between phases occurs directly at these shared boundary faces: velocity boundary conditions are enforced on the fluid side, while the computed pressure field applies forces to the solid surfaces.
This shared discretization enables bidirectional two-way coupling without introducing spurious leakage or artificial connectivity.

We validate our method on diverse scenarios including flows around thin membranes, geometries with narrow passages such as a stack of sheets and a valve, and deformable structures immersed in liquid.
These examples demonstrate the method's ability to handle arbitrarily thin structures, maintain precise leakproofness without artificial sealing, and capture complex fluid–structure interactions that were previously error-prone, as existing methods cannot robustly preserve correct topological connectivity of the fluid domain at finite resolution.

\vspace {10em}

\subsection{Contributions}
In summary, our work makes the following contributions:
\begin{itemize}
  \item \textit{Topology- and geometry-exact fluid discretization:}
  a flux-form finite-volume discretization on a modified Voronoi diagram that exactly enforces the path connectivity of the fluid domain induced by arbitrarily thin and codimensional solid geometry, preventing leakage without artificially sealing valid flow paths.

  \item \textit{Resolution-independent preservation of flow connectivity:} 
  a clipped Voronoi stitching procedure that reassigns orphaned cells to restore valid fluid connectivity when particle sampling is insufficient or locally absent, while maintaining a fixed number of fluid degrees of freedom. Valid fluid paths are preserved even under extreme under-resolution, including cases with no particles initially present in narrow passages, due to connectivity-aware mesh construction.
  
  \item \textit{Explicit bidirectional coupling on a conforming unstructured mesh:}
  a coupling formulation that uses the pressure Lagrange multiplier to both enforce incompressibility and transfer forces to solids, supporting a broad range of fluid–structure interaction scenarios on a shared discretization.

  \item \textit{Consistent pressure gauge handling under dynamic topology changes:}
  a practical strategy for maintaining pressure gauge consistency across dynamically separating and reconnecting connected components of the fluid domain, avoiding inconsistencies arising from independent pressure nullspaces.
\end{itemize}
\section{Related Work}

\begin{table*}[]
\caption{Summary of related works and their limitations as applied to our problem of bidirectional coupling of incompressible fluids with thin shells.}
\label{tbl:related_works_summary}
\begin{tabular}{lllll}
\hline
Method
& \multicolumn{1}{l}{%
\begin{tabular}[t]{@{}l@{}}
Solid Boundary \\
Treatment
\end{tabular}}
& Codimensional?
& \multicolumn{1}{l}{%
\begin{tabular}[t]{@{}l@{}}
Bidirectional \\
Coupling?
\end{tabular}}
& Failure Mode
\\ \hline

\rowcolor[HTML]{ECF4FF}
\multicolumn{1}{l}{%
\begin{tabular}[t]{@{}l@{}}
\citet{numerow2024differentiable} \\
(Differentiable Voronoi)
\end{tabular}}
& Clipped Voronoi & \xmark & \cmark & Volumetric, Sealing \\

\multicolumn{1}{l}{%
\begin{tabular}[t]{@{}l@{}}
\citet{de2015power} \\
(Power Particles)
\end{tabular}}
& Clipped Voronoi / Particles Sampling & \xmark & \cmark & Volumetric, Sealing \\

\rowcolor[HTML]{ECF4FF}
\multicolumn{1}{l}{%
\begin{tabular}[t]{@{}l@{}}
\citet{tao2022vempic} \\
(VEMPIC)
\end{tabular}}
& Cut Cells & \cmark & \xmark & Exploding DOFs, No Coupling \\

\multicolumn{1}{l}{%
\begin{tabular}[t]{@{}l@{}}
\citet{azevedo2016preserving}
\end{tabular}}
& Cut Cells & \cmark & \xmark & Exploding DOFs, No Coupling \\

\rowcolor[HTML]{ECF4FF}
\multicolumn{1}{l}{%
\begin{tabular}[t]{@{}l@{}}
\citet{peskin1972flow} \\
(Immersed Boundary Method)
\end{tabular}}
& Smoothed Delta Function & $\mathbf{\sim}$ (Volumized) & \cmark & Leaking \\ \hline
\end{tabular}
\end{table*}

\subsection{Solid-Fluid Coupling}

The computer graphics community has explored solid-fluid coupling across different solid and fluid discretizations, including Eulerian approaches \cite{teng2016eulerian}, Lagrangian methods \cite{akinci2012versatile, xie2023contact}, and hybrid techniques \cite{guendelman2005coupling, fang2020iq}.
The variational formulation of \citet{batty2007fast} has enabled robust bidirectional interaction, later extended by \citet{takahashi2020monolith} for true monolithic solid-fluid coupling.
Discretization limitations, particularly the mismatch between solid and fluid phases, have prevented their use on codimensional geometries.
A critical challenge lies in correctly identifying which portions of the fluid domain can interact with one another while respecting solid interfaces.
Ensuring that fluid interactions properly honor these topological constraints has remained a persistent difficulty.

By treating the fluid as a power diagram generated by Langrangian source points, \citet{de2015power} simulate the fluid phase using a discretization similar to the Voronoi-based approach we consider.
They, however, do not consider two-way coupling, and do not support thin solids as it requires a notion of volumes to resolve all its geometry.

Employing an absorbent fabric model, \citet{fei2018multi} investigated codimensional solid-fluid interaction, but their approach sidestepped leakproofness requirements entirely since the material model inherently allowed fluid penetration.

\citet{azevedo2016preserving} developed a cut-cell method for topology-preserving fluid simulation that maintains connectivity around solid obstacles.
Their approach clips grid cells at solid boundaries and introduces new degrees of freedom for each disconnected cut-cell region.
While this enables accurate geometric representation, the method suffers from a scalability limitation: as multiple thin solids stack within a single grid cell, each clipped region generates additional degrees of freedom.
Consequently, the total number of degrees of freedom can grow large, compromising computational efficiency.
Furthermore, their work addresses only one-way coupling, with the fluid responding to solids but not vice versa.
Similarly, \citet{tao2022vempic} tackled fluid simulation around arbitrary non-manifold solids with codimensional features, demonstrating capability for complex geometric configurations.
However, their method is constrained by computational requirements imposed by the explosion of degrees of freedom from their robust cut-cell mesh generator \cite{tao2019mandoline}, rendering it impractical for scenes with complex interactions.
In contrast, rather than increase the complexity of the fluid discretization as a time varying function fo the geometric and topological complexity of the thin solid, our method maintains the total number of degrees-of-freedom in the simulation as an invariant dictated by the number of fluid particles, avoiding this explosion of variables.
Our method achieves exact geometric matching with solid boundaries through a substantially simpler algorithm that avoids the computational overhead of complex cut-cell reconstruction.

While coupling of thin interfaces to fluids has been relatively underexplored in computer graphics, fluid-structure interaction (FSI) has received extensive attention in computational mechanics.
Three established classes of methods dominate this literature: arbitrary Lagrangian-Eulerian (ALE) interfaces \cite{donea1982arbitrary}, the ghost fluid method \cite{fedkiw1999non}, and the immersed boundary method \cite{peskin1972flow}, each offering distinct advantages but facing particular challenges with thin deformable structures.

Arbitrary Lagrangian-Eulerian methods employ a deforming fluid mesh that adapts to match the solid's mesh representation while maintaining a fixed Eulerian discretization in regions far from the interface \cite{donea1982arbitrary}.
The primary challenge lies in preserving mesh quality during deformation, which constrains the method to small displacements and volumetric solid representations.
While recent extensions have explored auxiliary coarse meshes to handle larger deformations with codimensional solids in incompressible flows \cite{fernandes2019ale}, these advances remain restricted to 2D scenarios and relatively simple geometries.

The ghost fluid method originated as a technique for sharp interface treatment in Eulerian multiphase simulations \cite{fedkiw1999non}.
SPH methods have adapted analogous techniques, populating solid regions with fictitious particles to mitigate kernel support deficiencies \cite{takeda1994numerical, randles1996smoothed, schechter2012ghost}.
However, the smoothing inherent to SPH kernels inherently blurs interface sharpness, and the technique necessarily treats solids volumetrically.
A fundamental challenge persists across these implementations: misalignment between the actual solid boundary geometry and the boundaries of fluid discretization elements.
Our method circumvents this issue entirely by aligning the fluid mesh boundaries precisely with solid geometry, enabling straightforward enforcement of impermeability through velocity boundary conditions at solid faces.

\begin{figure}[t]
  \centering
  \includegraphics[width=\linewidth]{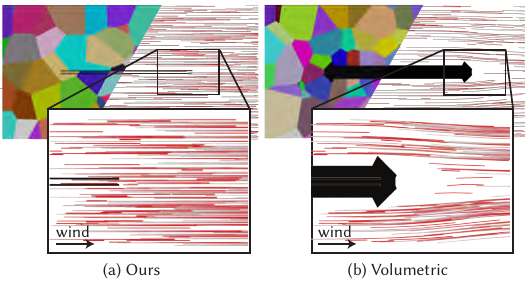}
  \caption{\textbf{Failure case of volumetrically resolving codimensional solids (b), compared to our approach (a).} A constant rightward velocity field is passed through two codimensional sheets. The Voronoi diagram (coloured cells) and streamlines (red lines) are shown. Notice that when volumetrically resolving the solid, the velocity field incorrectly bends around the solid as the expanded boundary condition now must be respected. The small channel inside is also completely blocked off. Our method preserves the small channel and properly resolves the velocity field.}
  \label{fig:r-flatsheet}
\end{figure}

\begin{figure}[t]
  \centering
  \includegraphics[width=0.7\linewidth]{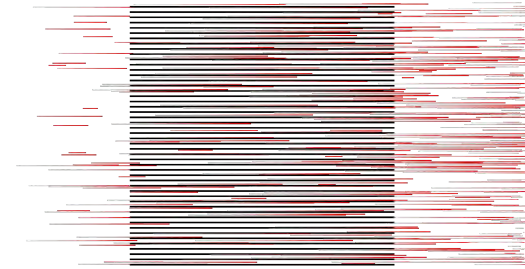}
  \caption{\textbf{Fluid pushing through a stack of codimensional thin geometry.} Our method successfully resolves even a challenging stack of many small orifices without clogging any.}
  \label{fig:r-flatsheet-2}
\end{figure}

The immersed boundary method places Lagrangian solid boundaries within an Eulerian fluid grid, employing smoothed delta functions to transfer forces from the Lagrangian onto the Eulerian fluid representation \cite{peskin1972flow, peskin2002immersed}.
The method suffers from leaking issues, with solid Lagrangian points able to traverse past fluid grid points during motion.
Additionally, force smoothing across a finite support region inherent to delta functions prevents exact velocity matching of fluid and solid phases, leading to fluid penetration.
Subsequent developments have pursued sharp boundary condition enforcement \cite{mittal2008versatile}, primarily through cut-cell and ghost cell methodologies that segment the Eulerian grid to constrain it to one side of the domain boundary.
\citet{ye1999accurate} developed a cut-cell technique that extends boundary cells to incorporate empty regions where grid points were removed by solid boundaries, deforming the regular grid into trapezoidal elements near boundaries.
Our stitching procedure shares conceptual similarities with this method, reassigning disconnected fluid regions to appropriate fluid source points.
We consider our approach a natural extension of cut-cell methods to unstructured Voronoi meshes, with the advantage of resolving subgrid-scale fluid pathways that previous techniques cannot capture.

\citet{numerow2024differentiable} presents a method for for coupling cell-based mechanics by representing each cell as a Voronoi region and computing differentiable gradients of cell volumes and interface areas, enabling gradient-based optimization and simulation of forces between cells.
This approach discretizes solids as volumetric clipping regions, leading to artifacting shown on Figure \ref{fig:cc-voro}, where empty cells are created if the geometry is too thin.

This limitation of volumetrically resolving solids, persistent among methods lacking sharp boundary discretization, detaches the codimensional boundaries away from their true spatial location.
This not only causes spurious forces, but also potentially destroys the actual topology of the fluid domain, with small orifices no longer becoming flowable, as shown on Figure \ref{fig:r-flatsheet}.
By comparison, our method cleanly resolves codimensional geometry, resolving the rightward flow even under multiple of these extremely small channels, shown on Figure \ref{fig:r-flatsheet-2}.

\subsection{Modified Voronoi Diagrams}

\begin{figure}[t]
  \centering
  \includegraphics[width=\linewidth]{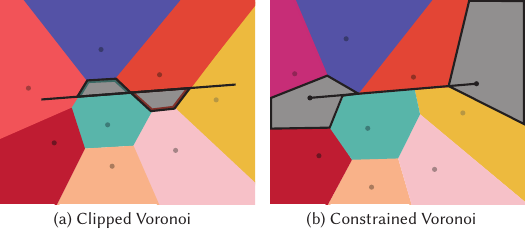}
  \caption{\textbf{Existing Voronoi meshing methods fail to mesh an immersed codimensional sheet.} (a) Clipped Voronoi methods use the codimensional sheet as a clipping plane, effectively deleting parts of the cell opposite of a sheet as if it were a halfplane.  (b) Constrained Voronoi methods treat vertices of the occluder as Voronoi sites, creating extra sites that are not allocated to any fluid particle. Notably, the example shown separates the domain into two separate fluid regions, breaking the original topology of the fluid domain.}
  \label{fig:cc-voro}
\end{figure}

Existing Voronoi methods are insufficient for simulating codimensional geometry as they fail to completely resolve the solid geometry at its true positions, as shown on Figure \ref{fig:cc-voro}.
Voronoi clipping methods treat solids as clipping halfplanes, deleting the geometry opposite of the source site \cite{yan2010efficient}.
This leaves voids in the fluid geometry that either must be treated as volumetric solids, thereby incorrectly expanding the solid surface, or treat it as a void that the fluid must now fill, forming a spurious divergence sink in the fluid domain.
Our method resolves this error by reattaching these voids into neighbouring sites, thereby including them once again into the fluid domain.

Constrained and visibility-aware Voronoi methods generalize standard Voronoi diagrams to account for geometric constraints or occlusions.
Classical constrained Voronoi constructions, as the dual of constrained Delaunay triangulations, introduce the constraint vertices as Voronoi sites \cite{tournois20102d, seidel1988constrained}.
This effectively gives sites the properties of the solid, expanding its representation volumetrically and potentially displacing boundary conditions and breaking the original topology of the fluid domain (Figure~\ref{fig:r-flatsheet}). 
Similarly, visibility-constrained Voronoi diagrams studied in computational geometry account for occlusion or directional visibility restrictions \cite{aurenhammer2014visibility, okabe2000spatial}, producing cells that may be non-convex, disconnected, and of quadratic combinatorial complexity.
In both families of methods, the presence of Voronoi sites at solid vertices or obstacle endpoints inherently introduces artificial thickness and modifies fluid-domain connectivity.

\citet{tsin1996geodesic} proposed a Voronoi construction where barriers constrain visibility without generating their own sites, but their formulation handles only rectilinear barriers in two dimensions.
This is much closer to our addressed scenario of codimensional solid boundaries that cannot generate their own volumes, but extending this work to arbitrary barrier geometries and three-dimensional domains presents significant theoretical and computational challenges and have yet to be addressed in the literature.
As such no existing Voronoi meshing algorithm is suitable for simulation without corrupting the corrupting the intended fluid domain topology.

In comparison to these methods which add sources to boundary sheets, we instead construct the Voronoi diagram exclusively from fluid particle sources, then explicitly clip the resulting cells to incorporate solid geometry.
Orphaned cells generated by this clipping are reassigned through a stitching procedure that preserves path connectivity.
We effectively solve the voiding issue of standard clipped Voronoi algorithms.
This strategy directly embeds solid interfaces into the fluid partition, achieving the sharp boundary resolution and guaranteed fluid topology preservation that visibility-based approaches lack.

\section{Equations of Motion}

Let $\Omega$ denote the fluid domain, and $\partial\Omega$ represent the solid boundary.
This boundary may consist of arbitrarily thin structures, including codimensional surfaces that partition the fluid domain.
We aim to solve incompressible fluid flow within $\Omega$, with velocities matching the solid velocities at $\partial\Omega$.

The fluid motion within $\Omega$ is governed by the incompressible Euler equations:
\begin{align}
    \frac{\partial \mathbf{u}}{\partial t} + (\mathbf{u} \cdot \nabla)\mathbf{u} + \frac{1}{\rho}\nabla p &= \mathbf{f}, \label{eq:incompressible_momentum} \\
    \nabla \cdot \mathbf{u} &= 0, \label{eq:incompressible_continuity}
\end{align}
where $\mathbf{u}(\mathbf{x},t)$ is the fluid velocity, $p(\mathbf{x},t)$ is the pressure, $\rho$ is the constant fluid density, and $\mathbf{f}$ represents body forces such as gravity. At the solid boundary $\partial\Omega$, we enforce velocity boundary conditions:
\begin{align}
    \mathbf{u} \cdot \hat{\mathbf{n}} = \mathbf{u}_s \cdot \hat{\mathbf{n}} \quad \text{on} \quad \partial\Omega, \label{eq:boundary_condition}
\end{align}
where $\mathbf{u}_s$ is the prescribed velocity of the solid surface.
This condition ensures no fluid penetration through solid boundaries while allowing the fluid to slip tangentially along moving surfaces.
The boundary condition couples the fluid and solid phases.
The solid boundary adds the velocity boundary condition to the fluid, and the pressure Lagrange multiplier acts to both enforce incompressibility, as well as apply a contact force onto the solid, enabling bidirectional coupling.

\section{Topology-Preserving Discretization}

Arbitrarily shaped solids impose boundary conditions that our discretization must resolve, including thin shells and other codimensional interfaces.
Fixed-resolution methods and volumetric solid representations demand excessive refinement to capture such thin structures and are thus unappealing as a discretization.
Static meshes and deformable mesh methods with fixed connectivity such as ALE are also unappealing, being typically limited to low-deformation volumetric solids, with issues arising when trying to resolve boundary fluids next to the solid mesh under mesh distortion or inversions \cite{fernandes2019ale, shamanskiy2021mesh}.

Lagrangian fluid representations naturally satisfy these requirements.
We adopt the Voronoi-based discretizations of \citet{borgers2005lagrangian}, where Lagrangian particles track fluid motion and induce a spatial partition, assigning each particle the region of space closest to it.

\subsection{Challenges to Leakproofing}

The Voronoi mesh induced by the set of fluid points does not automatically match the solid boundaries. 
In order to have sharp resolution of the solids, care must be taken to make sure that the faces of our fluid discretization to match the faces of the solids.
Proper coupling of boundary conditions with fluid particles necessitates modifying the Voronoi diagram to incorporate boundaries as faces.
Furthermore, a particle's domain on one side of a barrier must not extend past that barrier unless a contiguous fluid path exists to that side.
In other words, each particle's domain must represent a single contiguous fluid region.

Prior methods that enforce solid boundaries volumetrically such as \citet{de2015power} cannot resolve arbitrarily thin features exactly. 
The ``thickening'' approach proves poorly suited for immersed thin structures, as demonstrated in Figure \ref{fig:r-flatsheet}, where two flat sheets is immersed in a tangentially flowing fluid.
Inviscid fluid flow parallel to the thin solid surface should not induce any change in the fluid velocity.
However, the volumetric approach necessitates a finite thickness, causing the fluid to need to curve around the thickened solid.
This ``thickening'' approach is also poorly suited for flow through small orifices, with the gap between the two sheets becoming completely closed.
Such numerical clogging of small holes is problematic as it can cause artifacting in simulation of heart valves for example.
Our method guarantees flow anywhere the fluid connectivity allows for, avoiding such clogging behaviour.

\begin{figure}[t]
  \centering
  \begin{subfigure}[t]{0.49\linewidth}
    \centering
      \includegraphics[width=\linewidth]{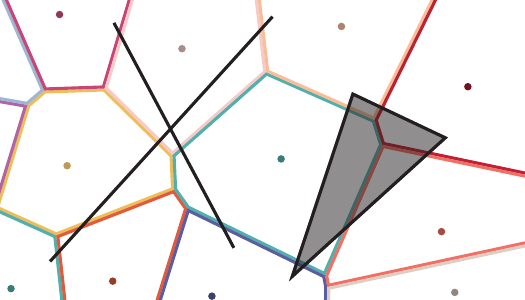}
    \caption{Initial Voronoi diagram (coloured outlines), clipped by solid constraints (black outlines, black shaded).}
    \label{fig:meshing-1}
  \end{subfigure}
  \begin{subfigure}[t]{0.49\linewidth}
    \centering
      \includegraphics[width=\linewidth]{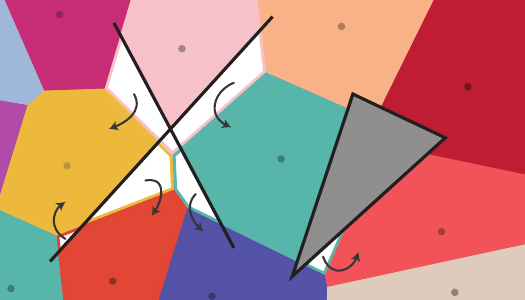}
    \caption{Cells containing a source point are assigned (filled), while the rest are registered as ``orphaned'' (white). Orphan cells will inherit the valid neighbours with the shortest proxy distance (grey arrow).}
    \label{fig:meshing-2}
  \end{subfigure}
  \begin{subfigure}[t]{0.49\linewidth}
    \centering
      \includegraphics[width=\linewidth]{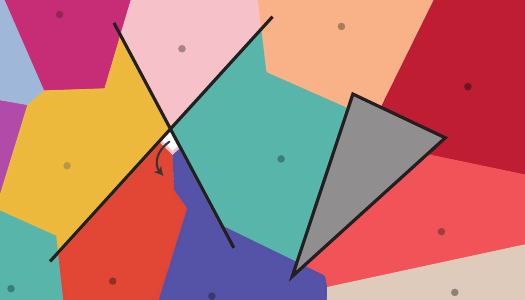}
    \caption{New connectivity reveals more orphaned cells that can be inherited.}
    \label{fig:meshing-3}
  \end{subfigure}
  \begin{subfigure}[t]{0.49\linewidth}
    \centering
      \includegraphics[width=\linewidth]{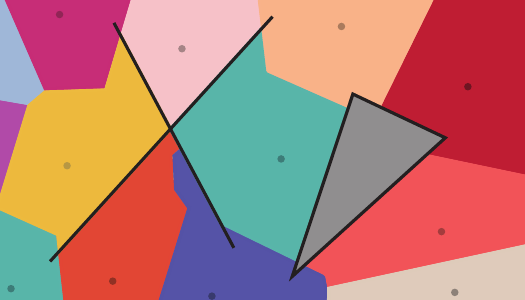}
    \caption{Final modified Voronoi diagram.}
    \label{fig:meshing-4}
  \end{subfigure}
  
  \caption{Stitching orphaned cells back to valid cells from the clipped Voronoi based on path-connectivity constraints.} \label{fig:methods-voronoi}
  \Description{Stitching orphaned cells back to valid cells from the clipped Voronoi based on path-connectivity constraints.}
\end{figure}

We thus seek a discretization that exactly recovers the connectivity produced by solid boundaries, is leakproof where and only where required, and permits fluid flow wherever a valid path exists.

\begin{algorithm}[t]
\caption{Clipped Voronoi Stitching}
\label{alg:meshing}
\begin{algorithmic}[1]
\State Compute initial Voronoi tessellation
\State Clip Voronoi cells by solid boundaries
\For{each cell}
    \If{cell does not contain its generating point}
        \State Label cell as \texttt{orphaned}
    \EndIf
\EndFor
\While{any cell is orphaned}
    \For{each orphaned cell}
        \For{each neighboring non-orphaned cell}
            \State Compute path length from cell centroid, to face centroid, to neighbour's cell centroid
        \EndFor
        \State Merge orphaned cell with neighboring non-orphaned cell with shortest path length
        \State Update orphaned cell label
    \EndFor
\EndWhile
\end{algorithmic}
\end{algorithm}

\subsection{Clipped Voronoi Stitching Algorithm}

We present a stitching algorithm that embeds solid surfaces directly into the Voronoi mesh, introducing new interfaces that subdivide cells into disjoint fluid components when solids obstruct connectivity. Patches of space are reassigned to appropriate fluid particles in a path-connected manner, producing a mesh whose topology correctly reflects solid-induced separation, consequently enabling fluid to continue flowing wherever a valid path exists around the solid but only along those valid paths.

Our method is initialized by constructing the standard Voronoi diagram induced by the fluid particles over all of space. 
All solid faces are then included in this structure by clipping all Voronoi cells with these faces.
Voronoi faces inside volumetric solids are removed, while codimensional solid faces are added as new faces in the diagram.
A neighbourhood graph is constructed by tagging each fluid face with the label of bordering fluid cells, where each edge represents a valid fluid-only path through space.

\begin{wrapfigure}{r}{1.5in}
  \centering
  \vspace{-0.5em} 
  \hspace{-2em}
  \includegraphics[width=\linewidth]{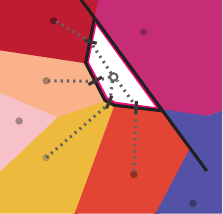}
  \label{fig:methods-paths}
  \vspace{-1.5em}
\end{wrapfigure}

A number of clipped cells around these solid faces will become ``orphaned'', meaning they no longer contain a source point within the cell.
These cells are tagged, then each orphaned cell is assigned to one of its non-orphaned neighbours, as shown in Figure \ref{fig:methods-voronoi}.
We choose to attached cells greedily, based on which neighbour has the shortest path from the orphan's centroid, to the shared face centroid, to the neighbour's site, as shown on the inset above.

This is essentially an easy to compute proxy for a geodesic distance lying solely within the fluid domain (\emph{i.e.} does not cross solids).
While a graph solve would be more accurate in giving the true shortest paths than the greedy approach, in practice most orphan cells are only 1 jump away from a valid cell, as shown on Table \ref{tbl:jumps}.
Multiple jumps appear for thin, poorly resolved channels between solid boundaries where the solid already significantly constrains the neighbourhood paths (see the maze in Section \ref{sec:maze})
This procedure is performed iteratively, as some orphaned cells may be landlocked by other orphaned cells until those cells are assigned.

\begin{table}[]
\caption{Number of jumps required to reach a valid cell on Frame 1. Initially valid cells are indicated with 0 jumps.} \label{tbl:jumps}
\begin{tabular}{lllll}
\hline
                      & 0      & 1     & 2   & 3+   \\ \hline
\rowcolor[HTML]{ECF4FF} 
Bunny                 & 7480 & 836 & 32 & 248 \\
Maze (0 particles)    & 6357  & 9   & 3  & 130 \\
\rowcolor[HTML]{ECF4FF} 
Maze (Fully resolved) & 9156 & 528  & 4  & 0    \\
Fan                   & 8216 & 75  & 0   & 0    \\
\rowcolor[HTML]{ECF4FF} 
Valves                & 14316 & 197  & 0   & 0    \\
Parachute & 1440 & 14 & 0 & 0 \\\hline
\end{tabular}
\end{table}

The aim of attaching cells through shortest geodesic distance is to keep good mesh quality, with the target being as convex as possible while respecting the flow connectivity.
Convexity is widely recognized in the literature as a desirable property for mesh elements, as it promotes numerical stability and well-conditioned discretizations for elliptic problems such as the pressure Poisson equation underlying projection methods \cite{batdorf1997computational, katz2011mesh}.
Near-convex, well-shaped cells produce uniform stencil coefficients, reduce skewness, and help avoid spurious pressure modes.
Poorly shaped or highly non-convex cells can cause large variations in the discrete operator, slowing solver convergence and producing inaccurate or oscillatory pressures.
Maintaining high-quality convex cells on unstructured meshes thus improves both solver robustness and the fidelity of the computed pressure, especially near complex solid boundaries.

Algorithm \ref{alg:meshing} presents this stitching process.
Our orphan-cell reassignment conceptually resembles breadth-first region growing or flood fill, but is adapted to prioritize the shortest proxy geodesic distance, and operate on clipped Voronoi cells.

\subsection{Pressure Projection}

Now that we have constructed our discretization, we seek to solve the usual pressure projection problem given by the momentum equation,
\begin{align}
    \frac{\rho}{\Delta t}\mathbf{u} + \nabla p = \frac{\rho}{\Delta t}\mathbf{u}^*, \label{eq:linear_system_1}
\end{align}
and divergence-free constraint,
\begin{align}
    -\nabla\cdot p = 0 \label{eq:linear_system_2}
\end{align}
with velocities $\mathbf{u}$ and pressures $p$. 

We aim to use the usual Chorin projection \cite{chorin1968numerical} to solve this system:
\begin{align}
    \Delta p &= \frac{\rho}{\Delta t} \nabla\cdot \mathbf{u}^* \label{eq:poisson}\\
    \mathbf{u} &= \mathbf{u}^* - \frac{\Delta t}{\rho}\nabla p. \label{eq:velocity_update}
\end{align}

What remains then is the relevant discretization of each of the above operators suitable for our mesh. 
We take pressures and velocities to be collocated, meaning that the discrete Laplacian $L$ is simply the area-weighted graph Laplacian of the neighbourhood connectivity, with entries:
\begin{align}
    L_{ij} &= \frac{A_{ij}}{l_{ij}V_i} \\
    L_{ii} &= -\sum_{j} L_{ij}
\end{align}
\begin{wrapfigure}{r}{1.2in}
  \centering
  \vspace{-0.5em} 
  \hspace{-2em}
  \includegraphics[width=\linewidth]{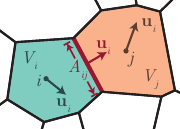}
  \label{fig:methods-local_voronoi}
  \vspace{-1.5em}
\end{wrapfigure}

where $j\in \mathcal{N}(i)$ are neighbours of $i$, $l_{ij}=||\mathbf{x}_j-\mathbf{x}_i||_2$ is the distance between cell sites, and $A_{ij}$ is the length of the face between $i$ and $j$.
While collocated pressures and velocities are problematic for regular grid regimes, we do not encounter the checkerboarding nullspace for our pressure solve as the connectivity graph between cell centers are unstructured in our case. 
It is vanishingly unlikely that every path pointing back to the query cell takes an odd number of jumps in the neighbourhood graph. 
Should this be a concern, implementing Rhie-Chow interpolation can alleviate this issue \cite{rhie1983numerical}, but from our experimentation it was unnecessary.

Like in a regular grid, we evaluate the right-hand side of Equation \ref{eq:poisson} by applying the divergence theorem and integrating the contribution of each face:
\begin{align} \label{eq;discrete-divergence}
    \left. \frac{\rho}{\Delta t} D\mathbf{u}^* \right|_i &= \frac{\rho}{\Delta t\ V_i} \left[ \sum_j (\mathbf{u}_{ij}\cdot\hat{\mathbf{n}}_{ij})a_{ij} + \sum_b (\mathbf{u}_{b}\cdot\hat{\mathbf{n}}_{b})a_{ib} \right],
\end{align}
where $j\in \mathcal{N}(i)$ and $b$ are elements of the solid boundary neighbouring $i$, with normals $\hat{\mathbf{n}}_{b}$ (outwards-facing from the fluid side).
$\hat{\mathbf{n}}_{ij}$ are the outward facing normals for fluid-fluid faces from $i$'s perspective.
With velocities being collocated, we evaluate face velocities by taking a simple average $\mathbf{u}_{ij} = \frac{1}{2}(\mathbf{u}_i + \mathbf{u}_j)$.
The corresponding gradient operator is as follows:
\begin{align}
    \left.Gp\right|_i = \frac{1}{V_i}\sum_j a_{ij}(p_j-p_i)\hat{n}_{ij}
\end{align}

\section{Discrete Gauss' Theorem}
\label{sec:discrete_flux_conservation}
\begin{wrapfigure}{r}{1.5in}
  \centering
  \vspace{-0.5em} 
  \hspace{-2em}
  \includegraphics[width=\linewidth]{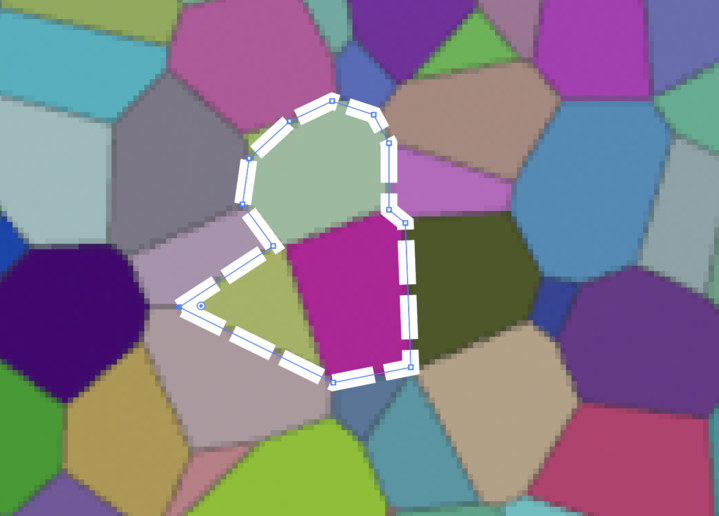}
  \vspace{-1em}
\end{wrapfigure}

Since our divergence operator is defined in flux form over Voronoi 
control volumes (Eq.~\ref{eq;discrete-divergence}), it satisfies a discrete Gauss' theorem. 
When summing discrete divergence over any union of Voronoi cells, fluxes across shared faces appear with opposite orientation and cancel exactly. As a result, the total integrated divergence over the region equals the fluxes across the boundary of the region (Gauss' theorem).

The discrete Gauss' theorem implies that, after pressure projection, the total flux across the boundary of any union of Voronoi cells vanishes up to solver tolerance. 

\begin{figure}
    \centering
    \includegraphics[width=0.9\linewidth]{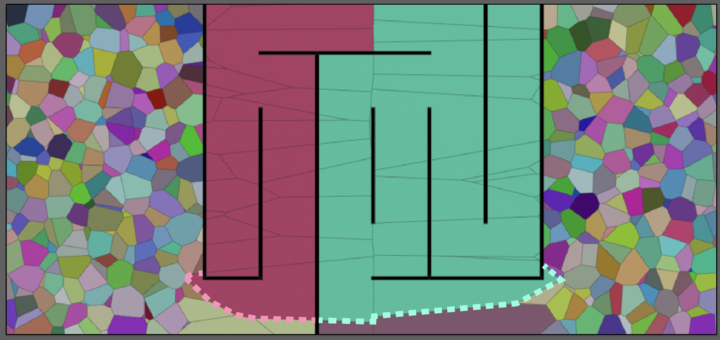}
    \caption{\textbf{Discrete resolution-independent flux cancellation.} In this configuration, the entire maze interior is represented by only two Voronoi cells (particles). Despite this coarse discretisation, summing the flux-form divergence over the path-connected maze region causes all internal face fluxes to cancel exactly, leaving only boundary fluxes at the two openings. After pressure projection, these boundary fluxes must balance, so the total flux entering the maze equals the total flux exiting it, ensuring exact flux balance between fluid regions connected only through the narrow maze.}
    \label{fig:v-maze}
\end{figure}

The maze depicted in Figure \ref{fig:r-maze} illustrates this how this discrete conservation property symbiotically interacts with guaranteed preservation of path-connectivity. The Voronoi stitching algorithm guarantees that the maze interior is a path-connected set of Voronoi cells whose boundary is formed by solid walls and two openings. Because divergence is discretised in flux form, all fluxes across internal faces of the maze cancel exactly when summed over the interior cells, as depicted in Figure \ref{fig:v-maze}. 

Consequently, the total flux entering the maze through one opening must equal the total flux exiting through the other opening, regardless of the complexity of the internal geometry or the number of fluid particles used in the discretization. 

\subsection{Gauge Pressure}

Incompressible simulation lacking pressure-Dirichlet boundary conditions only define pressure up to an additive constant. While in many standard fluid-only simulations this constant-pressure nullspace is inconsequential, careful handling becomes critical with the inclusion of dynamic solids that can alter the topology of the fluid domain. In particular, solids can partition a fluid into multiple disjoint regions, each of which could drift independently in pressure if the gauge is not consistently maintained.

Without proper management, the pressures on either side of a solid separating two disjoint fluid regions could shift by different constant offsets, leading to non-physical behavior such as sudden jumps in forces applied to the solid. Interestingly, this issue can arise even when iterative solvers are \emph{warm-started} from the previous timestep, since each disjoint region could still internally settle to a different constant pressures.

To address this, we explicitly maintain a consistent gauge across all fluid regions. At each timestep, we first identify disjoint fluid regions from the neighbourhood connectivity graph. For each disjoint region, we compute the volume-averaged pressure both from the previous timestep and the current raw solution. A constant shift is then applied to the pressure in that region so that the volume-averaged pressure matches the previous timestep. In cases where the total volume of the region has changed due to solid motion leading to expansion or compression, we also adjust the pressure according to the ideal gas law, $P_\text{new} V_\text{new} = P_\text{old} V_\text{old}$, to ensure physically consistent updates. This procedure ensures that the fluid on either side of a solid remains compatible, preventing spurious pressure offsets and improving the stability and realism of solid--fluid interactions.

\subsection{Timestepping}

Our proposed Voronoi-based solid–fluid coupling algorithm uses an explicit timestepping scheme for both the fluid and solid updates, shown in Algorithm \ref{alg:timestepping}.
At each timestep, we first handle the fluid phase, solving for a divergence free boundary condition that respects the solid velocity boundary conditions at its current state.

The pressure Lagrange multiplier is used to both enforce incompressibility in the fluid state, as well as is transferred to the solid state by integrating the pressure force at each solid face $F_s$:
\begin{align}
\left.\mathbf{F}_{fluid \to solid} \right|_{F_s} = - \sum_{j \in \mathcal{N}(F_s)} p_j A_{F_sj}\hat{\mathbf{n}}_{F_s j} \label{eq:fluidtosolid}
\end{align}
where $\mathcal{N}(F_s)$ is the set of fluid particles who share a face with $F_s$, and ${A}_{ij}$ is the area of the shared face with outwards-pointing normal $\hat{\mathbf{n}}_{F_s j}$.

\begin{algorithm}[H]
\caption{Voronoi-Based Solid–Fluid Coupling Timestep} \label{alg:timestepping}
\begin{algorithmic}[1]
\Require Fluid state $(\mathbf{x}_f^n, \mathbf{v}_f^n, p^n)$, Solid mesh $\mathcal{M}_s^n$ with vertex velocities $\mathbf{v}_s^n$ at time $t^n$
\Ensure Updated states $(\mathbf{x}_f^{n+1}, \mathbf{v}_f^{n+1}, p^{n+1})$, $(\mathcal{M}_s^{n+1}, \mathbf{v}_s^{n+1})$ at $t^{n+1} = t^n + \Delta t$

\State \textbf{Generate Mesh} (Algorithm \ref{alg:meshing})
\State \quad Construct a stitched Voronoi mesh using the current fluid positions $\mathbf{x}_f^n$ and solid mesh $\mathcal{M}_s^n$

\State \textbf{Fluid Evolution}
\State \quad Apply external forces to update fluid velocities $\mathbf{v}_f^n \to \mathbf{v}_f^*$
\State \quad Solve pressure projection to update pressure $p^n \to p^{*}$ (Eq. \ref{eq:poisson})
\State \quad For each connected fluid domain, match its gauge from the previous timestep $p^* \to p^{n+1}$
\State \quad Update fluid velocities $\mathbf{v}_f^{*} \to \mathbf{v}_f^{n+1}$ (Eq. \ref{eq:velocity_update})
\State \quad Advect fluid positions $\mathbf{x}_f^n \to \mathbf{x}_f^{n+1}$ using $\mathbf{v}_f^{n+1}$

\State \textbf{Fluid-to-Solid Force Coupling}
\State \quad Compute fluid forces on solid mesh from pressure field (Eq. \ref{eq:fluidtosolid})

\State \textbf{Solid Evolution}
\State \quad Update solid mesh $\mathcal{M}_s^{n+1}$ using off-the-shelf solvers (rigid body or FEM)
\end{algorithmic}
\end{algorithm}

With these forces, the solid update is handled by any standard off-the-shelf solver.
Our examples use both a rigid body and a deformable solid finite-element solver.
All state updates, including fluid velocity, fluid positions, and solid mesh configurations, are performed using forward Euler integration, making the overall method simple, efficient, and straightforward to implement while maintaining a physically consistent coupling between the fluid and the solid.

\subsection{Interpolation}

The Lagrangian fluid discretization may result in locally underresolved fluid regions.
These may potentially happen nearby to solid structures if the fluid enters a solid constriction where not many fluid particles are allowed to flow through. 
In these cases, however, the solid geometry provides additional information through known surface velocities, independent of the bulk fluid motion.
This enables the use of interpolation to enhance visual fidelity via marker particle advection.
We emphasize that this is strictly a postprocessing step: the true fluid particles have already had their physics computed during simulation based on the incompressible Navier-Stokes evolution described above.

We apply Shepard interpolation to approximate the velocity on a marker particle from the face velocities of the cell it resides in \cite{shepard1968two}.
This consists of finding the shortest distance from the query point to each face of the enclosing cell, and performing inverse-distance weighting on those face's velocities.
We note that there are no guarantees on the incompressibility of the interpolated field, though incompressible interpolations is a ripe avenue for research in recent years \cite{chang2022curl, nabizadeh2024coflip}, especially in our case where we aim to support nonconvex cells with codimensional obstacles.

\section{Results}

\begin{table*}[]
\caption{Scene parameters and runtimes of presented examples.} \label{tbl:metrics}
\begin{tabular}{lllllll}
\hline
\rowcolor[HTML]{FFFFFF} 
\textit{} & \multicolumn{4}{l}{\cellcolor[HTML]{FFFFFF}\textit{}}                  & \textit{Solid Treatment and} & \textit{}        \\
\rowcolor[HTML]{FFFFFF} 
\textit{} & \textit{Domain} & \textit{$\#$Frames} & \textit{Time} & \textit{$\#P$} & \textit{$\#V$}               & \textit{s/Frame} \\ \hline
\rowcolor[HTML]{ECF4FF} 
Bunny     & 1 x 1 x 1       & 180                 & 3             & 7480           & 6880 Static                  & 3.3              \\
\rowcolor[HTML]{FFFFFF} 
Fan       & 1.5 x 1.5 x 1.5 & 720                 & 12            & 8216           & 2405 Scripted                & 52.6             \\
\rowcolor[HTML]{ECF4FF} 
Maze      & 1 x 0.2 x 0.04  & 120                 & 1             & 6357 - 9156    & 88 Static                    & 70.5             \\
\rowcolor[HTML]{FFFFFF} 
Valve     & 3 x 1 x 0.125   & 2880                 & 24            & 14316          & 104 Deformable               & 26.06            \\
\rowcolor[HTML]{ECF4FF} 
Parachute      & 8 x 16 x 8  & 300                 & 5             & 1440    & 200 Deformable + 26 Rigid                    & 19.5             \\
\hline
\end{tabular}
\end{table*}

We implemented our method as a solver node in Houdini \cite{houdini}, and have attached code to the supplemental material.
Houdini’s built in rigid body and FEM solvers were used for solid simulation \cite{houdini}. 
Scene parameters are provided in Table \ref{tbl:metrics}.

\subsection{Sealed Bunny}

\begin{figure}[t]
  \centering
  \includegraphics[width=\linewidth]{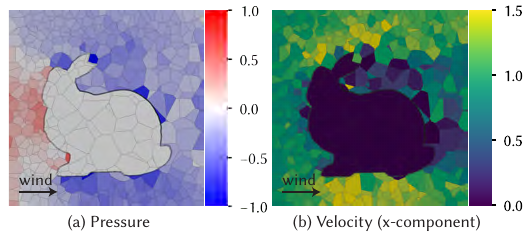}
  \caption{\textbf{Watertight sealed bunny inside a windtunnel.} Notice that the fluid in the interior remains perfectly quiescent, while the exterior air is pushed past the bunny by an inflow boundary to the left of the scene.}
  \label{fig:r-bunny}
\end{figure}

We place a static bunny inside a wind tunnel, shown on Figure \ref{fig:r-bunny}.
The exterior fluid is initialized with a velocity of $u_x=1$ lengthwise across the bunny, while the interior fluid is initialized with zero velocity.
Inflow and outflow boundaries are placed at either end of the wind tunnel to maintain the exterior fluid velocity.
We find that for the interior of the bunny, our leakproof treatment maintains the static fluid up to machine single-precision, demonstrating that our method is able to maintain separate uninteracting fluid domains through a thin shell.

As a stress test of our stitching algorithm, we also simulate this with the same exterior fluid particles, but seed only a single fluid particle on the bunny interior.
We find that the stitching algorithm is able to correctly assign all cells inside the bunny to that single particle.
This means the entire geometry is represented by a single particle's state.
This state remained fully watertight and experienced no fluid connectivity with the external fluid, thus retaining its original values up to machine precision.

\subsection{Maze} \label{sec:maze}

\begin{figure}[t]
  \centering
  \includegraphics[width=\linewidth]{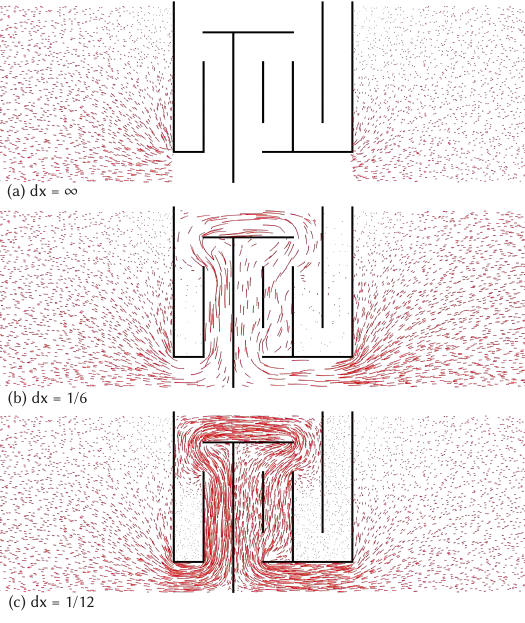}
  \caption{\textbf{Flow through a maze through different resolutions of the interior fluid.} An incompressible fluid passes between input and output reservoirs connected by a narrow, winding maze. We progressively reduce the fluid resolution inside the maze, including a case with no initial particles in the corridor. In all cases, the output velocity is preserved, demonstrating that our stitching method maintains flow connectivity and topology independently of resolution.}
  \label{fig:r-maze}
\end{figure}

To demonstrate the low-resolution capability of our stitching method, we run a fluid through a tight maze (Figure \ref{fig:r-maze}), consisting of an input and output reservoir, with codimensional walls creating a winding pathway between the two reservoirs.
We initialize the fluid in both reservoirs with an initial velocity of $u_x=1$, and apply a inflow and outflow boundaries to the two ends to maintain this fluid velocity.

We run an ablation of the same simulation setup at decreasing fluid resolutions within the maze, demonstrating that flow is still preserved through the maze irrespective of the resolution.
At worst case, we remove all initial particles within the maze, and show that a fluid path is still constructed between fluid degrees of freedom between the input and output reservoirs via our stitching method.
We thus show that our method guarantees that the fluid will always preserve the topology of the fluid region (\emph{i.e.} flow connectivity) regardless of resolution. 
Table \ref{tbl:jumps} shows that as the resolution of the interior is decreased, more jumps are required for the many interior orphaned cells to find a path to a valid fluid cell.

Because the fluid is incompressible, regardless of how the fluid moves within the maze, the flow must maintain this same velocity at the output reservoir.
As mentioned on Section \ref{sec:discrete_flux_conservation}, the Voronoi control volumes satisfies a discrete Gauss' theorem.
Two connected control volumes are shown on Figure \ref{fig:v-maze}.
As such, the fluid flowing into one side must necessarily be balanced by the same amount of fluid exiting the opposite side.
We verify that for any collection of cells, including cells entering into the maze, the total integrated flux over the boundary of their control volumes is zero up to machine precision.


\subsection{Fan}

\begin{figure}[t]
  \centering
  \includegraphics[width=\linewidth]{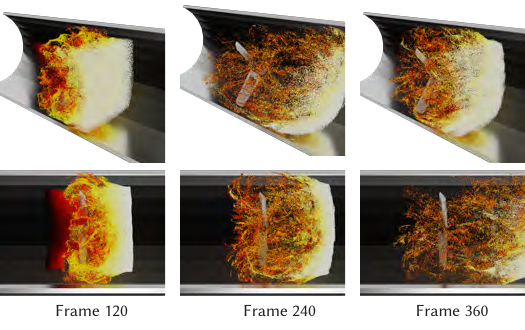}
  \caption{\textbf{Scripted fan forming vortices as it spins up.} Note that the blade geometry is significantly smaller than the particle resolution.}
  \label{fig:r-fan}
\end{figure}

We push air using a scripted rotating fan shown on Figure \ref{fig:r-fan}.
We note that this example relies heavily on the resolution of subgrid surfaces, as the fan is considerably thinner than the particle resolution.
Even here, our method is able to transfer velocity from the solid to the fluid phase, with the fan generating a toroid-shaped vortex as it spins up, pulling air from the outside edge of the fan and pushing it from the center.

\subsection{Flexible Valve}

\begin{figure}[t]
  \centering
  \includegraphics[width=\linewidth]{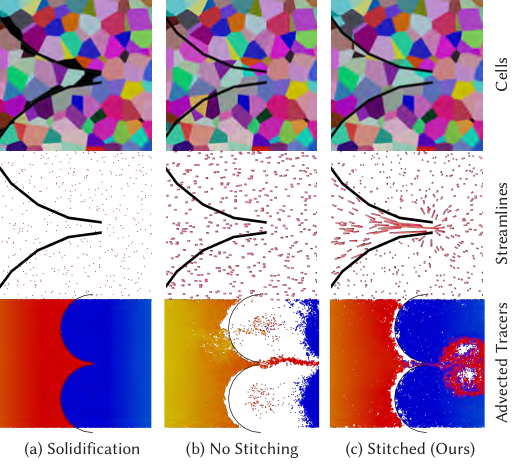}
  \caption{\textbf{Incorrect handling of orphaned cells destroys local fluid topology, resulting in improper simulation of the valve.} Zoom in of the area near the valve opening are shown in the top two rows, with (above) cells coloured randomly according to their assigned fluid particle IDs and (middle) particle velocities traced. (Below) shows the full simulation output at Frame 100 with advected tracers coloured based on their starting position relative to the valve. (a) shows the result of using normal clipped Voronoi cells, where orphaned cells are simply dropped and treated as solids. This expansion of the solid boundary away from its original spatial position is prone to becoming overly leakproof, sealing the small opening of the valve and causing no fluid to flow. (b) shows naively reattaching the orphans onto their original generating site. This causes a topological link across the boundary, which allows the velocity to flow across the boundary, generating leaking. Our method (c) resolves the solid boundary sharply and perfectly preserves fluid topology, treating the solid boundary as a perfect separator between distinct fluid regions.}
  \label{fig:r-valve-ids}
\end{figure}

We simulate the fluid around a flexible heart valve responding to a sinusoidal body force with a 3.5 second period, resulting in an average forward velocity of 0.25 on the push stroke (systole) and a backward velocity of -0.1 on the relaxation (diastole). 
The far ends of the tubes are given traction-free inlet boundary conditions, and are allowed to spawn particles when underresolved (when the number of particles within 0.25 units of the boundary fall below the starting resolution).

We run the valve under two different conditions: a healthy valve and a defective valve.
The defective valve is constructed by deleting the center tenth (0.5 units) of the geometry to form a larger opening, representative of a condition known as annular dilation \cite{hurst2017heart}.

We show that for the healthy valve, when the fluid flows in the direction of the valve, the fluid is able to open the valve walls and freely flow.
When the fluid flows in the opposite direction, the valve walls shut, producing a perfectly leakproof interface where the fluid on either side no longer interact.
Our method is able to handle coupling with a thin solid interface, and handles topology changes naturally with the separation and rejoining of the two fluid domains.

We see that in the defective valve, significant backflow is created, producing a qualitatively different result from a very minor change in input geometry.
The difference between the two valves hinges on the ability to differentiate geometries where the valve closes shut as opposed to allowing a small amount of flow through--a distinction that our method naturally captures by sharply resolving the solid surface.

\subsection{Parachute}

\begin{figure}[t]
  \centering
  \includegraphics[width=\linewidth]{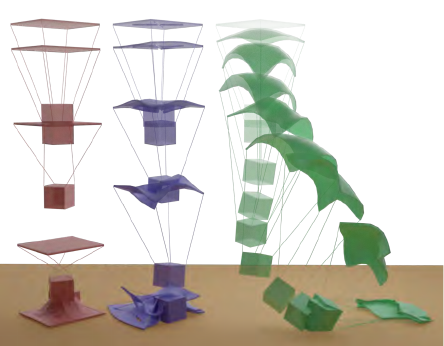}
  \caption{\textbf{Comparison of falling parachute simulations under different physical conditions, shown in 30 frame (0.5 s) intervals.} (green, right) A solid parachute simulated with our leakproof solid–fluid coupling method, exhibiting strong drag and a decelerated descent. (blue, center) A torn parachute simulated with the same leakproof method, allowing air leakage through the tears and resulting in reduced drag and faster descent. (red, left) Reference freefall in the absence of air, with no aerodynamic forces and purely ballistic motion under gravity.}
  \label{fig:r-parachute}
\end{figure}

We drop a rigid box attached to a deformable soft parachute with springs representing the rope in three separate scenarios: without our method, with our method on a torn parachute, and with our method on an intact parachute (Figure \ref{fig:r-parachute}).
Without our method, the box freefalls under gravity.
With our method, as the parachute falls, it encounters the still air, causing a local negative divergence, corresponding to a high pressure zone in the solve.
This pressure is then transferred to the parachute, counteracting its fall until the velocity of the solid and air match representing force balance.
This results in a slow steady descent of the object.

In the torn parachute, a thin x-shaped tear is introduced.
As it falls and the tear opens, a fluid path between the bottom and the top of the parachute is formed, allowing air pressure to equalize on either side of the fabric. This reduces the potential pressure gradient on either side of the sheet, thereby reducing its ability to induce drag and arrest the object's fall.
\section{Discussion and Conclusion}
We present a method for coupling incompressible fluids with solids while respecting fluid connectivity, creating a discretization that is leakproof only when required and flow-permissive otherwise.
Our method leverages the Voronoi diagram to partition space, modifying it to conform to solid boundaries and enabling natural force transfer between the solid and fluid regimes.

Several extensions of our work present promising avenues for future research.
Our method is significantly bottlenecked by the cost of computing the Voronoi diagram, as we currently recompute the Voronoi structure at each timestep, consuming $85\%$ of our simulation runtime.
Leveraging temporal coherence to update the previous timestep's structure offers a promising approach for acceleration \cite{guibas2018kinetic}.
However, we note that the diagram is only required for connectivity and face areas.
A method for approximating face areas while maintaining the same connectivity properties without fully computing a Voronoi diagram could significantly improve performance.

Additionally, our modifications to the Voronoi diagram sidesteps the true intended structure, which is a visibility-constrained Voronoi.
In such a structure, every cell should contain the domain of all points in space closest to it, while accounting for visibility constraints imposed by solid boundaries.
Conceptually, if source points were to grow in the domain given some occluding planes, the points would propagate waves that naturally bend around the ends of the occluding planes.
Some work exists for rectilinear occluders in 2D \cite{tsin1996geodesic}, but extensions to arbitrary barriers in 3D remain an open problem.
Similar approaches could also enable more accurate intracell gradients, as gradients to faces should vary according to occluders within these cells.

An exploration of divergence-free interpolation schemes that support nonconvex cell shapes and codimensional geometries would also be a significant boon, as we highlight that our method recovers the true flow connectivity of the fluid domain irrespective of resolution.
As such, being able to run at lower resolution is a key strength, and as such a more accurate divergence-free interpolation of velocities within poorly shaped cells would further enhance its visual appeal, producing smoother and more visually accurate flow even at coarser resolutions.

Finally, our choice of an explicit method means coupling is handled through multiple timestepping.
The fluid imposes boundary conditions on the solid, which then evolves and imposes boundary conditions back onto the fluid.
A more accurate approach would find the force balance between the two regimes \emph{within} each timestep, such as with a Newton iterative solver.
This would avoid potential timestepping constraints that lead to jittering behavior of the solid as the solid and fluid attempt to agree on the local pressure and velocities at the solid surface.

\bibliographystyle{ACM-Reference-Format}
\bibliography{bib}

\end{document}